\documentclass[a4paper,twocolumn,11pt,accepted=2023-10-30,showpacs,amsmath,amssymb,floatfix,superscriptaddress,notitlepage]{quantumarticle}
\pdfoutput=1
\usepackage[utf8]{inputenc}
\usepackage[numbers,sort&compress]{natbib}
\usepackage[left=2cm,right=2cm]{geometry}
\usepackage{physics}
\usepackage{amsmath}
\usepackage{amsfonts}
\usepackage{amssymb}
\usepackage{mathtools}
\usepackage{xcolor}
\usepackage{graphicx}
\usepackage{dsfont}
\usepackage{amsthm}
\usepackage[hidelinks]{hyperref}

\hypersetup{%
  colorlinks=true,
  linkcolor=blue,
  citecolor=blue,
  urlcolor=blue,
}

\begin{document}

\title{Bounding the Minimum Time of a Quantum Measurement}

\author{Nathan Shettell}
\affiliation{Centre  for  Quantum  Technologies,  National  University  of  Singapore,  Singapore 117543, Singapore}

\author{Federico Centrone}
\affiliation{ICFO-Institut de Ciencies Fotoniques, The Barcelona Institute of Science and Technology,
08860 Castelldefels (Barcelona), Spain}

\author{Luis Pedro Garc\'ia-Pintos}
\affiliation{Joint Center for Quantum Information and Computer Science and Joint Quantum Institute, University of Maryland, College Park, Maryland 20742, USA} 
\affiliation{Theoretical Division (T4), Los Alamos National Laboratory, Los Alamos, New Mexico 87545, USA}


\begin{abstract}
Measurements take a singular role in quantum theory.
While they are often idealized as an instantaneous process, this is in conflict with all other physical processes in nature. 
In this Letter, we adopt a standpoint where
the interaction with an environment is a crucial ingredient for the occurrence of a measurement. Within this framework, we derive lower bounds on the time needed for a measurement to occur.
Our bound scales proportionally to the change in entropy of the measured system, and decreases as the number of of possible measurement outcomes or the interaction strength driving the measurement increases.
We evaluate our bound in two examples where the environment is modelled by bosonic modes and the measurement apparatus is modelled by spins or bosons.
\end{abstract}

\maketitle

\section{Motivation}
\noindent Quantum measurements are one of the most controversial and enigmatic aspects of quantum theory. Although most agree on the end result of said measurements, the dynamics behind them has been a subject of debate since the initial `quantum-boom' in the early twentieth century~\cite{bohr1928quantum, wigner1984review, bub2010two, schlosshauer2013snapshot}. 
The textbook description of a
quantum measurement is the `collapse of the wave function'~\cite{heisenberg1949, stapp1972}, which was later codified by von Neumann in his mathematical formulation of quantum mechanics. In it, a quantum measurement is defined as a probabilistic, non-unitary, irreversible, and instantaneous process~\cite{von2018}. This is in direct contention with another postulate of quantum mechanics: that all microscopic processes are unitary and reversible. This peculiar exemption for measurements (along with other concerns) is dubbed the `measurement problem'~\cite{brukner2017quantum, zurek2003decoherence, zurek2009quantum, schlosshauer2005decoherence, schlosshauer2007decoherence}.

One of the most successful attempts to address the measurement problem without changing the fundamental postulates of quantum mechanics is the decoherence mechanism introduced by Zeh~\cite{zeh1970interpretation}. Decoherence can be viewed as 
the practically 
irreversible generation of correlations between a system of interest and a large environment.
Although the global evolution of the system and environment is unitary, the system 
follows a non-unitary dynamics that rapidly damps coherence within the system, destroying quantum interference effects, and driving the state to a statistical mixture that, for all practical purposes, can be described by a classical probability distribution~\cite{zurek2003decoherence, zurek2003decoherence,schlosshauer2005decoherence}. 
To what extent decoherence can solve the measurement problem is still matter of debate~\cite{joos1985,brukner2017quantum,schlosshauer2019quantum}.
However, it did have a huge impact on current interpretations of quantum mechanics and in the development of many modern branches of quantum theory,
motivating
carefully controlled experiments designed to track the dynamics of a quantum measurement~\cite{brune1996observing, jordan2010uncollapsing, minev2019catch, carlesso2022present}. 

In this Letter, we derive a general bound on the minimum time it takes for a measurement to occur in decoherence-based interpretations.
We stress that our results are not aimed at deciding whether a measurement has occurred or not. Instead, this Letter caters to those that follow the school of thought in which 
a measurement is facilitated by the interaction with an environment.
Alternatively, whatever philosophy the reader may have, the results we show imply bounds on the decoherence timescales of the correlations of arbitrarily complex quantum systems,
setting a saturable bound on the minimum timescales needed for a quantum system to effectively behave classically due to the interaction with an environment \cite{zurek2009quantum}.

\section{Measurement Model}
\begin{figure}[t]
    \centering
    \includegraphics[width=0.5\textwidth]{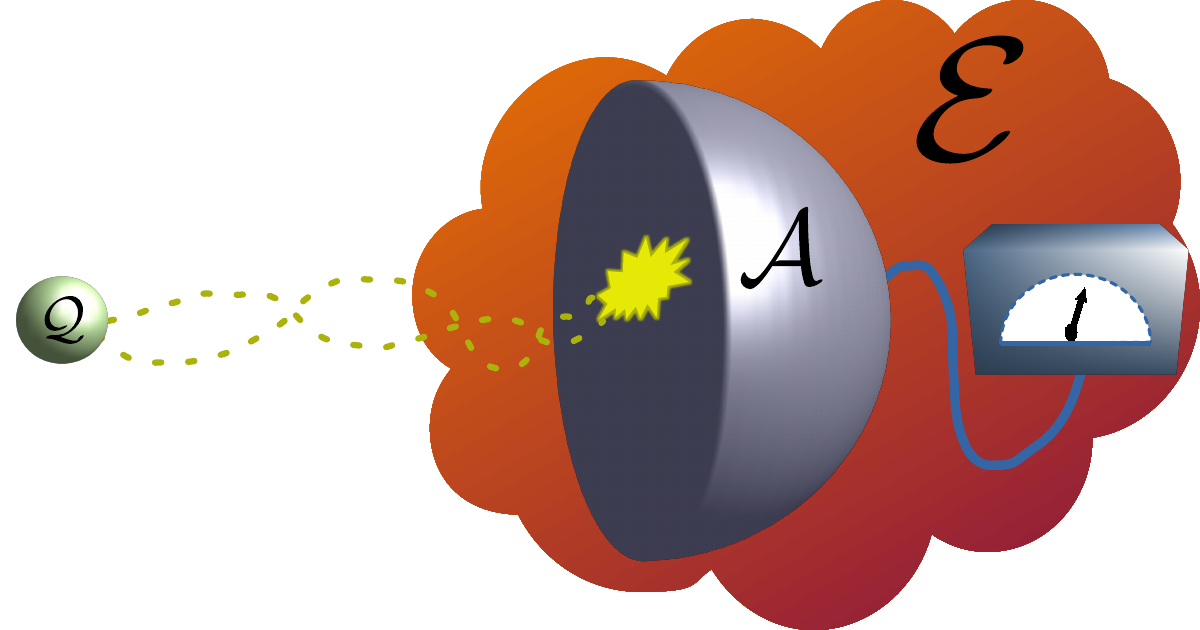}
    \caption{Our measurement scheme is composed of three parts: a system $\mathcal{Q}$ and a measurement apparatus $\mathcal{A}$ in contact with an environment $\mathcal{E}$.
    During the pre-measurement stage,  $\mathcal{A}$ interacts with the system $\mathcal{Q}$, correlating to the potential outcomes of a measurement. At this stage, the process is still reversible. However, the posterior interaction with an environment $\mathcal{E}$ drives the system and apparatus to a state physically indistinguishable from a statistical mixture of measurement outcomes, in a process that is practically irreversible. 
    }
    \label{fig:QAE}
\end{figure}

\noindent We consider a setup where  a quantum system $\mathcal{Q}$ interacts with a measurement apparatus $\mathcal{A}$ capable of recording the outcome of the measurement. Thereafter, 
an external environment $\mathcal{E}$ interacts with the measurement apparatus~\cite{breuer2002theory}. 
Note that the assumption of a mesoscopic measurement device ensures that outcomes are observable at a macroscopic level. The apparatus consists solely of the elements that interact with the measured system, e.g. the free charges in an avalanche photodiodes or the mean photon number in a superconducting microwave resonator. All the remaining constituents that would conventionally be considered  part of the device are treated as a thermal environment that conveys the  information related to the quantum state in the classical world.
In order to differentiate between a measurement and a noisy process, we assume that $\mathcal{Q}$ does not interact with $\mathcal{E}$, see Fig.~(\ref{fig:QAE}).

Within the ideal measurement scheme devised by Von Neumann \cite{von2018}, a quantum measurement is partitioned into two distinct transformations: a `pre-measurement' followed by a `collapse'. During the pre-measurement of an observable $O=\sum_j o_j \Pi_j^O$, the apparatus $\mathcal{A}$ and system $\mathcal{Q}$ interact in such a way that the `pointer states' $\ket*{a_j}$ of the apparatus, become correlated with the projectors $\Pi^{O}_j$~\cite{zurek1981pointer}. More precisely, if the system is initialized in $\ket*{\psi^\mathcal{Q}}$, 
the state after the pre-measurement stage is
\begin{equation}
    \label{eq:pre-measurement}
    \ket*{\psi^\mathcal{QA}} = \sum_j \big( \Pi^{O}_j \ket*{\psi^\mathcal{Q}} \big) \otimes \ket*{a_j}.
\end{equation}
Since $\ket*{\psi^\mathcal{QA}}$ is a pure state that can be undone by applying local unitaries on the joint system of $\mathcal{Q}$ and $\mathcal{A}$, the pre-measurement does not constitute a definite quantum measurement~\cite{schlosshauer2005decoherence}.
Moreover, without an additional physical process, one cannot unambiguously identify the outcomes of a measurement. The secondary phase of the measurement, the `collapse', resolves the ambiguity  by transforming $\ket*{\psi^\mathcal{QA}}$ into the statistical mixture
\begin{equation}
    \label{eq:collapse}
    \rho^\mathcal{QA}_\mathcal{M} = \sum_{j} \Pi^{O}_j \dyad*{\psi^\mathcal{Q}} \Pi^{O}_j \otimes \dyad*{a_j}.
\end{equation}
This state is physically indistinguishable from a classical probability distribution where outcome state $\dyad*{a_j}$, corresponding to the outcome $o_j$, occurs with probability $\Tr(\Pi^{O}_j \dyad*{\psi^\mathcal{Q}} \Pi^{O}_j)=\expval*{\Pi^{O}_j}{\psi^\mathcal{Q}}$. Although Eq.~\eqref{eq:collapse} is conventionally thought of as an instantaneous jump process~\cite{von2018}, decoherence theory introduces an environment $\mathcal{E}$ which allows modelling this transition as a continuous transformation on the joint state of $\mathcal{QAE}$.

A Hamiltonian which  governs the transition dynamics of $\ket*{\psi^\mathcal{QA}} \rightarrow \rho^\mathcal{QA}_\mathcal{M}$ can be expressed as a sum of terms
\begin{equation}
    \label{eq:Hamiltonian}
    H=H_\mathcal{QA}+H_\mathcal{E}+H_\mathcal{AE},
\end{equation}
where the subscripts indicates which system(s) a specific Hamiltonian affects, for example $H_\mathcal{E}$ acts exclusively on the environment. We assume that the Hamiltonian commutes with the pointer basis, $[H,\Pi^{O}_j \dyad*{a_j} \Pi^{O}_j] = 0$, which ensures that measurement outcomes in the pointer basis are stable under the evolution \cite{zurek2003decoherence,schlosshauer2005decoherence}. Moreover, given a large environment, the process $\ket*{\psi^\mathcal{QA}} \rightarrow \rho^\mathcal{QA}_\mathcal{M}$ typically occurs rapidly\footnote{One may be concerned by a `fuzzy' definition of measurement that relies on the state of the system merely becoming \emph{close} to $\rho^\mathcal{QA}_\mathcal{M}$. More definite, objective notions arise if quantum gravity implies fundamental uncertainties in measurements~\cite{GambiniLPPullin2019}}. Note that we have excluded interaction terms between $\mathcal{Q}$ and $\mathcal{E}$ in Eq.~\eqref{eq:Hamiltonian}, i.e. $H_\mathcal{QE}$ and $H_\mathcal{QAE}$, this is to distinguish a measurement from a noisy process. Thus the unique term driving the measurement is $H_\mathcal{AE}$. Nevertheless, the results which follow in the next sections are easily extended to Hamiltonians which contain an interaction term between $\mathcal{Q}$ and $\mathcal{E}$.

The exact amount of time it takes for the process $\ket*{\psi^\mathcal{QA}} \rightarrow \rho^\mathcal{QA}_\mathcal{M}$ to occur could  be computed by solving the appropriate master equation. However, doing so often relies on a series of assumptions. Further, this definition of measurement time can bring about ambiguity due to the fact that typical decoherence inducing Hamiltonians cause the state of $\mathcal{QA}$ to become arbitrarily close to $\rho^\mathcal{QA}_\mathcal{M}$, but not necessarily equal.
In what follows, we derive a lower bound on the minimum timescales needed for this `collapse' stage of the measurement process to occur, while circumventing solving the dynamics of a Master equation, and ambiguity of deciding `when' a measurement has concluded.
We accomplish this by applying techniques from quantum limits \cite{margolus1998maximum, taddei2013quantum,adcQSL2013, deffner2013quantum, garcia2022}, from which we obtain a bound which is solely dependent on the relative entropy between $\ket*{\psi^\mathcal{QA}}$ and $\rho^\mathcal{QA}$, the number of measurement outcomes, and the interaction Hamiltonian between the apparatus and the environment $H_\mathcal{AE}$. In a similar spirit, the authors of Ref.~\cite{strasberg2022long} derived bounds on the timescale of the pre-measurement stage using quantum speed limit techniques. However, 
Ref.~\cite{strasberg2022long} focuses on unitary dynamics, whereas we take an open-system approach, resulting in different bounds on the timescales of measurements (see~\cite{margolus1998maximum} and~\cite{adcQSL2013,taddei2013quantum,deffner2013quantum} respectively).

\section{Bounding the Timescale of a Measurement} 

Our bound is based on a simple premise: if a car travels at a maximum speed of $v_\text{max}$, then the time needed to travel a distance $d$ must be greater than $d/v_\text{max}$. 
In quantum theory, one can consider the speed of a state or of an operator instead of a car, and derive quantum speed limits~\cite{margolus1998maximum, taddei2013quantum, adcQSL2013, deffner2013quantum, garcia2022}.

We aim for a bound on the time it takes for the system and apparatus to reach the post-measurement state $\rho^\mathcal{QA}_\mathcal{M}$. For that purpose, we focus on the speed of the relative entropy $S \big(\rho^\mathcal{QA} (t) \big\| \rho^\mathcal{QA}_\mathcal{M} \big) \coloneqq  \Tr \rho^\mathcal{QA}(t) \big( \ln \rho^\mathcal{QA}(t)-\ln \rho^\mathcal{QA}_\mathcal{M} \big)$ between the joint state of the system and apparatus at a given time, $\rho^\mathcal{QA}(t)$, and the post-measurement state, $\rho^\mathcal{QA}_\mathcal{M}$. Here, 
 $S(t) \nobreak \coloneqq \nobreak  -\Tr \rho^\mathcal{QA}(t) \ln \rho^\mathcal{QA}(t)$ is the entropy of $\rho^\mathcal{QA}(t)$. The relative entropy serves as a proxy of distance between two states, characterizing the probability to distinguish them \cite{Vedral2002RelativeEntropy}. In a hypothesis testing scenario, the probability to confuse two states $\rho_1$ and $\rho_2$ after $K$ measurements satisfies $p_\textnormal{error} \sim e^{-K \cdot S(\rho_1 \| \rho_2)}$~\cite{hiai1991proper}.

Substituting relative entropy in place of distance in the aforementioned car analogy, we arrive at a precursor for the bound on the measurement time
\begin{equation}
\label{eq:precursorbound}
    \tau \geq \frac{\delta S}{\max \left| \tfrac{d }{dt} S(\rho^\mathcal{QA} (t) \big\| \rho^\mathcal{QA}_\mathcal{M}) \right|},
\end{equation}
where $\delta S = S \big(\dyad*{\psi^\mathcal{QA}} \big\| \rho^\mathcal{QA}_\mathcal{M} \big)$ is the change in relative entropy. Note that
\begin{equation}
    S\left( \rho^\mathcal{QA}(t) \big\| \rho^\mathcal{QA}_\mathcal{M} \right) = -S(t) - \Tr \rho^\mathcal{QA} (t) \ln \rho^\mathcal{QA}_\mathcal{M},
\end{equation}
and since the post-measurement state is diagonal in the pointer basis, so is the operator $\ln \rho^\mathcal{QA}_\mathcal{M}$. Therefore,
\begin{equation}
\begin{split}
    &\Tr \frac{d \rho^\mathcal{QA}(t)}{dt} \ln \rho^\mathcal{QA}_\mathcal{M} \\
    =& -\frac{i}{\hbar} \Tr \Big( \Big[ H, \rho^\mathcal{QAE} (t) \Big] \ln \rho^\mathcal{QA}_\mathcal{M}  \otimes \mathds{1}_\mathcal{E} \Big)  \\
    =& -\frac{i}{\hbar} \Tr \Big( \Big[\ln \rho^\mathcal{QA}_\mathcal{M}  \otimes \mathds{1}_\mathcal{E}, H \Big]\rho^\mathcal{QAE} (t) \Big) \\
     =& 0,
\end{split}
\end{equation}
where the final equality makes use of the fact that $\Pi^{O}_j \dyad*{a_j} \Pi^{O}_j$ commutes with the total Hamiltonian. Hence, it is enough to focus on the change in entropy $S(t)$ of combined state of $\mathcal{Q}$ and $\mathcal{A}$ to compute the bound expressed in Eq.~\eqref{eq:precursorbound}. However, determining the maximum of $\big| \frac{dS(t)}{dt} \big|$ often requires solving complex dynamics. In order to circumvent this challenging task, we make use of the inequalities
\begin{equation}
\label{eq:ineq1}
    \hbar \left| \frac{dS(t)}{dt} \right| \leq \hbar \, \Delta S \sqrt{I_d} \leq 2 \Delta S \Delta H_\mathcal{AE},
\end{equation}
proven in Ref.~\cite{garcia2022}.
Here,
\begin{equation}
    I_d = \sum_j \frac{1}{\lambda_j} \left( \frac{\partial \lambda_j}{\partial t} \right)^2
\end{equation}
is the incoherent portion of the quantum Fisher information of $\rho^\mathcal{QA}(t)$ with eigenvalues $\{ \lambda_j \}$,
\begin{equation}
    \label{eq:varentropy}
    \big(\Delta S \big)^2 = \Tr \big( \rho^\mathcal{QA}(t) \ln^2 \rho^\mathcal{QA} (t) \big)-S(t)^2
\end{equation}
is the varentropy of state $\rho^\mathcal{QA}(t)$,
 and
\begin{equation}
    \big(\Delta H_\mathcal{AE} \big)^2=\Tr \big( \rho^\mathcal{QAE} (t) H_\mathcal{AE}^2 \big)- \big[ \Tr \big( \rho^\mathcal{QAE} (t) H_\mathcal{AE} \big) \big]^2
\end{equation}
is the variance of the interaction Hamiltonian $H_\mathcal{AE}$\footnote{While alternative bounds on the entropy rate have been derived~\cite{SIEPRA2007,deffner2021energetic,Mohan_2022}, the main advantage of Eq.~\eqref{eq:ineq1} is that it involves standard deviations instead of operator norms, which typically results in tighter bounds~\cite{garcia2022}.}. The quantities $I_d$, $\Delta S$ and $\Delta H_\mathcal{AE}$ are time-dependent, but their dependence is not made explicit in Eq.~\eqref{eq:ineq1}, nor subsequent equations, for the sake of clarity. The final inequality we employ is a bound on $\Delta S$ with respect to the number of non-zero eigenvalues of $\rho^\mathcal{QA} (t)$~\cite{reeb2015tight}, which is dictated by the number of distinct outcomes $A \nobreak = \nobreak \big| \{ \ket*{a_j} \} \big|$ that the apparatus can measure,
\begin{equation}
\label{eq:ineq2}
    \Delta S \leq \sqrt{\frac{1}{4}\ln ^2 (A-1) +1}=f_A.
\end{equation}
A proof can be found in the Appendix.
Note that $f_A = 1$ for a measurement with a binary output (such as the spin of a qubit), and $f_A \approx \frac{1}{2}\ln A$ for a measurement with $A \gg 1$.

Combining Eqs.~\eqref{eq:precursorbound}, \eqref{eq:ineq1} and \eqref{eq:ineq2}, we obtain bounds on the minimum timescale of a measurement
\begin{equation}
    \label{eq:timescaleGEN}
    \tau \geq \frac{\delta S}{\max \big( \Delta S \sqrt{I_d} \big)} \geq \frac{\hbar \, \delta S}{2 f_A \cdot \max \big( \Delta H_\mathcal{AE} \big)}.
\end{equation}
The tighter bound in Eq.~\eqref{eq:timescaleGEN} requires the information-theoretic quantity $ \Delta S \, \sqrt{I_d}$, typically hard to calculate,  
whereas the looser bound only requires knowledge of the uncertainty in the interaction Hamiltonian. 

The right-most bound in Eq.~\eqref{eq:timescaleGEN} can be further simplified in the strong-coupling regime \cite{casanova2010deep}, where $H_\mathcal{AE}$ dominates the evolution and is time-independent, since in this situation $\Delta H_\mathcal{AE}$ is (approximately) time-independent. Then, we obtain the main result of this Letter,
\begin{equation}
    \label{eq:timescale}
    \tau \geq \frac{\hbar \, \delta S}{2 f_A \Delta H_\mathcal{AE}}.
\end{equation}
Henceforth, we assume the strong-coupling regime and a time-independent Hamiltonian $H_\mathcal{AE}$ so that the right hand side of the bound solely involves time-independent quantities. In this way, our result provides a simple-to-evaluate expression for the minimum time needed for a measurement to occur.

Note how the quantities that define the lower bound are naturally linked to the problem at hand.
The further away the initial state is to the post-measurement state, as quantified by $\delta S$, the larger the lower bound on $\tau$ is. 
In contrast, a stronger coupling to the environment, as quantified by $\Delta H_\text{int}$, leads to potentially faster measurement processes.

\begin{figure}
    \centering
    \includegraphics[width=0.48\textwidth]{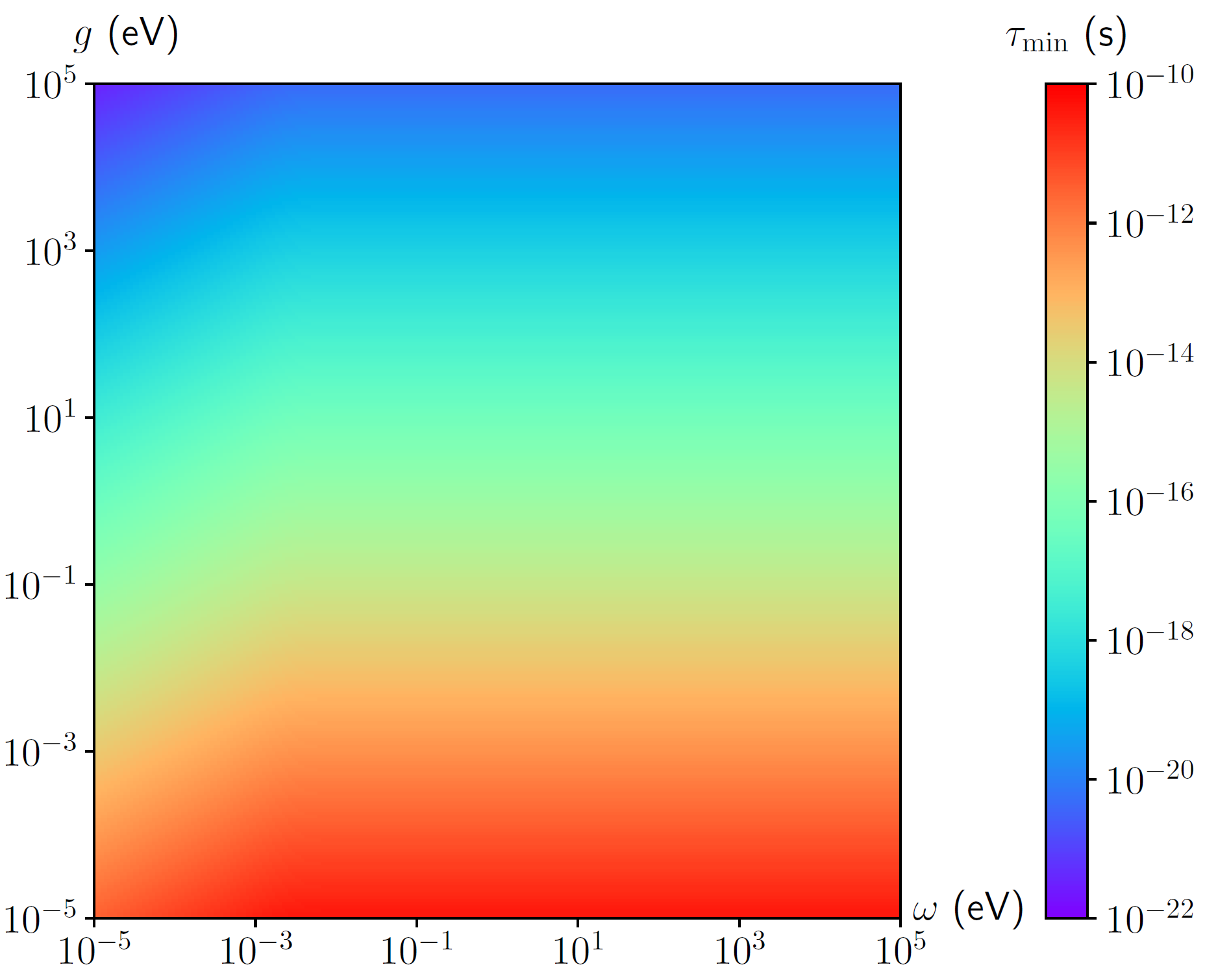}
    \caption{Plot of $\tau_\text{min}=\frac{\hbar \ln 2}{2 g}\sqrt{\tanh (\beta \omega/2)}$ ($N=1$) for a single environmental mode with coupling constant $g$ at $T=2$mK. Current attophysics techniques might allow to unravel the dynamics of a `slow' measurement up to $10^{-18}$s~\cite{gaumnitz2017streaking}.
    }
    \label{fig:plot}
\end{figure}


\ 
\section{Measurement times in the spin-boson model}
\noindent Let us consider the spin-boson, often used as a toy model to study environmental decoherence~\cite{leggett1987dynamics,schlosshauer2005decoherence, schlosshauer2007decoherence}. In it, spins representing the apparatus interact with a bosonic environment represented by a collection of harmonic oscillators in a state
\begin{equation}
    \label{eq:thermalenv}
    \rho^\mathcal{E} = \bigotimes_k \frac{1}{Z_k} e^{-\beta \omega_k a^\dagger_k a_k }.
\end{equation}
The subscript $k$ spans through the environmental modes, $\omega_k$ is the energy of the $k$th mode, $\beta = 1/(k_B T)$ is the inverse temperature, and $Z_k$ are normalization constants. The environment is assumed to be large such that $\rho^\mathcal{QAE}(t)=\rho^{QA}(t) \otimes \rho^\mathcal{E}$.

We assume $\mathcal{Q}$ to be a two-level system in an initial state $\ket*{\psi^\mathcal{Q}} = x \ket*{0} + y \ket*{1}$, 
and that the state of $\mathcal{QA}$
after the pre-measurement
is
\begin{equation}
\label{eq:GHZ}
    \ket*{\psi^\mathcal{QA}} = x \ket*{0}\ket*{ \downarrow }^{\otimes N} + y \ket*{1} \ket*{\uparrow}^{\otimes N}.
\end{equation}
Here, we assume that the apparatus is composed of $N$ distinct spin systems ($N \approx 1$ corresponds to microscopic apparatus, and $N \gg 1$ corresponds to a macroscopic apparatus). If we are interested in measuring the state of the system $\mathcal{Q}$ in the computational basis $\{\ket*{0},\ket*{1}\}$, the post-measurement state will be
\begin{equation}
    \rho^\mathcal{QA}_\mathcal{M} = |x|^2 \dyad*{0}\otimes\dyad*{\downarrow}^{\otimes N} + |y|^2 \dyad*{1}\otimes \dyad*{\uparrow}^{\otimes N}.
\end{equation}
Finally, the apparatus and environment interaction is modeled via the Hamiltonian~\cite{leggett1987dynamics, breuer2002theory}
\begin{equation}
    H_\mathcal{AE}^\text{SB} = \sum_{i=1}^N \sum_k \sigma_Z^{(i)} \otimes g_k ( a^\dagger_k +  a_k),
\end{equation}
where the local Hamiltonian $\sigma_Z^{(i)}=\dyad*{\downarrow}-\dyad*{\uparrow}$ acts as such exclusively on the $i$th spin and identity on all the others, the ladder operators $a_k$ and $a^{\dagger}_k$ act on the $k$th environmental modes degrees of freedom and $g_k$ is a positive coupling constant.

It holds that
\begin{equation}
\begin{split}
    (\Delta H_\mathcal{AE}^\text{SB} )^2 &= N^2 \sum_k g_k^2 \coth (\beta \omega_k/2) \\
    &= N^2 \int_0^\infty \hspace{-7pt} J(\omega) \coth(\beta \omega /2) \mathrm{d} \omega,
\end{split}
\end{equation}
where $J(\omega)$ is the spectral density of the coupling constants~\cite{schlosshauer2005decoherence}. Then, Eq.~\eqref{eq:timescale} implies that 
the measurement time satisfies 
\begin{equation}
\label{eq:sb-singlequbit}
    \tau_\text{SB} \geq \frac{\hbar \, \delta S}{2N} \left( \int_0^\infty \hspace{-7pt} J(\omega) \coth(\beta \omega /2) \mathrm{d} \omega \right)^{-1/2},
\end{equation}
with $\delta S = -|x|^2 \ln |x|^2-|y|^2 \ln |y|^2 \leq \ln 2$. 
Decreasing the coupling constants $g_k$ or the `size' $N$ of the measurement apparatus leads to necessarily larger measurement times.

The bound is illustrated in Fig.~(\ref{fig:plot}) with a microscopic apparatus ($N=1$) and a single environmental mode, in which the predominant order of magnitude is in the range of nanoseconds up to zeptoseconds. Current technologies might be able to witness a carefully engineered measurement, which in principle could be used as a test-bed to benchmark against objective collapse theories~\cite{marshall2003towards, kanari2021can, carlesso2022present} or fundamentally non-unitary dynamics~\cite{Penrose,Gambini,Blencowe}.

Even though the bound Eq.~\eqref{eq:timescale} is extremely general, Eq.~\eqref{eq:sb-singlequbit} correctly captures the timescales on the spin-boson model for a small apparatus. 
By solving the master equation in the Born-Markov approximation, one finds that the off-diagonal terms of the state acquire an exponentially decaying factor $r(t)=e^{- \Gamma}$, with~\cite{schlosshauer2007decoherence}
\begin{align}
    \label{eq:expdecay}
    \Gamma = 4 N \int_0^\infty \hspace{-4pt} \frac{J(\omega)}{\omega^2}(1-\cos (\omega t/\hbar) ) \coth (\beta \omega /2) \mathrm{d} \omega.
\end{align}
With this, we can find the time it takes for the relative entropy to fall to $\varepsilon \ll 1$. In an experiment,  $\varepsilon$ would be set, for instance, as a function of the precision with which the state can be estimated from tomography. Assuming $|x|=|y|=1/\sqrt{2}$ for simplicity, we get
\begin{align}
    2 \varepsilon &= (1+e^{-\Gamma} ) \ln \left(1+e^{-\Gamma}\right) + \left( 1-e^{-\Gamma} \right) \ln \left( 1-e^{-\Gamma} \right).
\end{align}
which simplifies to $\varepsilon \approx e^{-2\Gamma}$ by using that $\ln(1 \pm e^{-\Gamma}) \approx \pm e^{-\Gamma}$. Assuming that the measurement timescale is much shorter than the dynamical timescales imposed by the environmental modes, $t \ll \hbar/\omega_k$, we obtain the analytic solution 
\begin{equation}
\label{eq:tsolution}
    t \approx \frac{\hbar \sqrt{ \ln (1 / \varepsilon)}  }{2   \sqrt{N}} \Big( \int_0^\infty \hspace{-7pt} J(\omega) \coth(\beta \omega /2) \mathrm{d} \omega \Big)^{-1/2}.
\end{equation}

This expression is remarkably similar to the lower bound in Eq.~\eqref{eq:sb-singlequbit}, with the exception of the scaling in $N$. 
Notably, the Born-Markov approximation allows one to maximize $\Delta S \sqrt{I_d}$ and thus make use of the tighter bound in Eq.~\eqref{eq:timescaleGEN}. By doing so, one recovers the $\sqrt{N}$ scaling
\begin{equation}
    \label{eq:sb-timescale2}
    \tau_\text{SB} \geq \frac{\hbar \, \delta S}{2\sqrt{N}} \left( \int_0^\infty \hspace{-7pt} J(\omega) \coth(\beta \omega /2) \mathrm{d} \omega \right)^{-1/2}.
\end{equation}
A derivation is included in the Appendix.

\section{Measurement times in the boson-boson model}
\noindent Here, we consider a bosonic apparatus interacting with a bosonic environment. 
This was the theoretical framework used in \cite{brune1996observing}. The apparatus is represented by a cat state with pointer states $\ket*{\alpha}$ and $\ket*{-\alpha}$, where $|\alpha|^2$ is the mean number of excitations. At $t=0$ the state of $\mathcal{QA}$ is
\begin{equation}
    \label{eq:cat}
    \ket*{\psi^\mathcal{QA}} = x \ket*{0} \ket*{\alpha}+y\ket*{1} \ket*{- \alpha},
\end{equation}
while the post-measurement state is
\begin{equation}
    \rho^\mathcal{QA}_\mathcal{M} = |x|^2 \dyad*{0}\otimes \dyad*{\alpha} + |y|^2 \dyad*{1} \otimes\dyad*{-\alpha}.
\end{equation}
We label the creation and annihilation operators of the environment and the apparatus by $\{a_k,a_k^\dagger\}$ and $\{b,b^\dagger\}$, respectively.

The interaction Hamiltonian
\begin{equation}
    H_\mathcal{AE}^\text{BB} = \sum_k g_k (b a^\dagger_k +  b^\dagger a_k),
\end{equation}
results in $\{\ket*{\alpha},\ket*{-\alpha}\}$ as approximate pointer states~\cite{walls1985analysis,brune1992manipulation}.
Once more, we assume that the environment is in thermal equilibrium, Eq.~\eqref{eq:thermalenv}. In this example, when $|\alpha | \gg 1$, we find that
\begin{equation}
\begin{split}
    (\Delta H_\mathcal{AE}^\text{BB} )^2 &= |\alpha|^2 \sum_k g_k^2 \coth (\beta \omega_k/2) \\
    &= |\alpha|^2 \int_0^\infty \hspace{-7pt} J(\omega) \coth(\beta \omega /2) \mathrm{d} \omega.
\end{split}
\end{equation}
Then, the measurement time satisfies
\begin{equation}
\label{eq:bb-singlequbit}
    \tau_\text{BB} \geq \frac{\hbar \delta S}{2 |\alpha| \cdot \max (\Delta S)} \Big( \int_0^\infty \hspace{-7pt} J(\omega) \coth(\beta \omega /2) \mathrm{d} \omega \Big)^{-1/2}.
\end{equation}
We have explicitly left the varentropy dependence in the denominator, as opposed to using the inequality in Eq.~\eqref{eq:ineq2},  because the pointer basis only commutes with the Hamiltonian approximately. Moreover, $\ket*{\alpha}$ and $\ket*{-\alpha}$ are not exactly orthogonal.
In any case, upon solving the master equation and using the Born-Markov approximation, one finds these corrections to be negligible~\cite{walls1985analysis, breuer2002theory}, so that under a good approximation it holds that $\Delta S \lesssim f_A$\footnote{Alternatively, one could have chosen an alternative $H_\mathcal{AE}^\text{BB}$ to avoid the commutativity issue, e.g. $H_\mathcal{AE}^\text{BB} = b^\dagger b \sum_k g_k(a_k^\dagger + a_k)$~\cite{walls1985analysis}, however said Hamiltonian is representative of coupling the Fock states to the environmental modes, which is unrealistic and thus not typically used.}. 

Then, the bound is very similar to the spin-boson example, Eq.~\eqref{eq:sb-singlequbit}, so one can refer to Fig.~(\ref{fig:plot}) for insight on the dependencies with $g$ for a single environmental mode. The most noteworthy difference is the scaling of $1/|\alpha|=1/\sqrt{M}$, where $M$ is the average number of excitations, which differs with the scaling of $1/N$ in the spin-boson model\footnote{The scaling of $1/|\alpha|$ in our bounds seemingly disagrees with the one found in Refs.~\cite{brune1992manipulation,brune1996observing}, where they found a decoherence time that scales as $1/|\alpha|^2$. The difference is due to the different choice of interaction Hamiltonian~\cite{brune1992manipulation}.}
This model can be generalized to include more measurement outcomes, $A > 2$, by assigning additional phases to a coherent state for the additional measurement outcomes. For example if $A=4$ one could use $\ket*{\alpha}$, $\ket*{i\alpha}$, $\ket*{-\alpha}$ and $\ket*{-i\alpha}$~\cite{vlastakis2013deterministically}.

\section{Discussion} 

\noindent There is an uncomfortable dissonance between continuous physical processes in nature and the often-assumed instantaneous jump of a state during a quantum measurement. One proposed resolution relies on the fact that the interaction with an environment drives a measurement apparatus to a state that is, in practice,
indistinguishable from a statistical mixture of definite measurement outcomes~\cite{schlosshauer2005decoherence}.

Within this framework, we derived a general
bound on the minimum timescale of a quantum measurement. The bound is based on the principle that a measurement involves a change in entropy $\delta S$ in the system and measurement apparatus, 
and that this change takes a finite amount of time~\cite{garcia2022}. 
Crucially, the bound on the minimum measurement time, which depends on the energy variance of the apparatus-environment coupling, can be calculated without needing to solve for the exact complex dynamics induced by the environment. When the dynamics is simple enough, e.g., as in the spin-boson model with the Born-Markov assumption, the tighter bound Eq.~\eqref{eq:sb-timescale2} can be obtained by exploiting the information-theoretic bound in Eq.~\eqref{eq:timescaleGEN}.

While we focused on the time needed for an environment to drive the system and apparatus to the post-measurement state $\rho_{\mathcal{M}}^\mathcal{QA}$, an expression like Eq.~\eqref{eq:timescale} also bounds the time needed for a correlated quantum system to decohere. 

Bounds on the timescale of a quantum measurement have also been considered recently in Refs.~\cite{pokorny2020tracking, strasberg2022long, hu2022scalable}. However, the bounds in the literature assume particular models of open quantum dynamics or require solving the associated master equation for the density matrix of the system of interest. In contrast, we derive bounds that rely on few assumptions on the measurement model. 
We showcased our bound in two simple models. 
Remarkably, under the right regime, the bounds on the measurement times are within experimental reach, as illustrated in Fig.~(\ref{fig:plot}). It would be fascinating to devise an experiment, perhaps similar to~\cite{brune1996observing}, to probe these fundamental bounds.
This may require exploring more realistic measurement models.

Note that Eq.~\eqref{eq:timescale} is reminiscent of the Bremmerman-Bekenstein bound, which provides an upper limit on the rate at which information can be transferred given an energy constrain~\cite{BekensteinBound}. The original derivation of this bound is based on heuristic arguments, but more rigorous versions of it have been explored~\cite{garcia2022,deffner2010generalized}. Our bound provides a formal version that applies to the dynamics during a quantum measurement, providing new insights into the physics of information processing.
Finally, it would be interesting to study the relationship between our bound on the timescale of a measurement and the energetic constraints and resource costs of performing a measurement~\cite{jacobs2012quantum,NavascuesPRL2014, deffner2016quantum, guryanova2020ideal}.

\vspace{4pt}

\noindent
\emph{Acknowledgements. —} F.C. acknowledges the support of Severo Ochoa CEX2019-000910-S [MCIN/ AEI/10.13039/501100011033], Fundació Cellex, Fundació Mir-Puig, and Generalitat de Catalunya through CERCA.
L.P.G.P.~acknowledges funding by the DoE ASCR Quantum Testbed
Pathfinder program (award No.~DE-SC0019040), ARO MURI, DoE QSA, NSF
QLCI (award No.~OMA-2120757), DoE ASCR Accelerated Research in Quantum
Computing program (award No.~DE-SC0020312), NSF PFCQC program, AFOSR, AFOSR MURI, and DARPA SAVaNT ADVENT.
L.P.G.P.'s~work at Los Alamos National Laboratory was carried out under the auspices of the US DOE and NNSA under contract No.~DEAC52-06NA25396.

\bibliographystyle{plainnat}
\bibliography{main.bib}

\widetext
\clearpage
\appendix


\appendix

\section{Bound on the varentropy}

\setcounter{equation}{0}
\renewcommand{\theequation}{A\arabic{equation}}

In this appendix we report the proof of $\Delta S \leq f_A$, as taken from \cite{reeb2015tight}.  Note that the entropy, $S(\rho)$, and varentropy, $(\Delta S)^2$, depend both only on the non-zero eigenvalues of the density matrix $\rho$. In \cite{reeb2015tight}, the bound is given as a function of the dimension of the system, assuming that all eigenvalues of the density matrix are non-null and are distributed according to
\begin{align}
\label{formofrhowidehatinvariancethm}
{\rm spec}[\rho]~=~\left(1-r_d,\frac{r_d}{d-1},\ldots,\frac{r_d}{d-1}\right)~,
\end{align}
where $d$ is the dimension of $\rho$ and $r_d$ is a constant which maximizes the varentropy (and is dependent on $d$).

In our case, the bound is given as a function of the number of possible outcomes of the meter $\mathcal{A}$. This is because, although the dimension of the Hilbert space associated to the measurement apparatus can be significantly larger than $A$, the interaction with the environment preserves the non-vanishing eigenvalues. This is due to the definition of the pointer basis $\ket*{a_j}$, as
\begin{equation}\label{conditionPointer}
    [\Pi^{O}_j\dyad*{a_j}\Pi^{O}_j, H]=0.
\end{equation}
Note that in the strong-coupling regime this simplifies to $[\Pi^{O}_j\dyad*{a_j}\Pi^{O}_j, H_\mathcal{AE}]=0$.

It then follows that we can bound the varentropy using a bound with respect to a Hilbert space of dimension $d=A$. We include a proof for completeness.


\begin{proof}
For fixed $A\geq2$, we maximize the expression  of the varentropy over all probability distributions $\{p_i\}$ (i.e., the non-null spectra of $\rho$), which leads to the Lagrange function
\begin{align}\label{lagrangefunctionforvariance}
L(\{p_i\},\nu)~:=~\sum_i p_i(\ln p_i)^2-\big( \sum_i p_i\ln p_i \big)^2\,+\,\nu\sum_i p_i~,
\end{align}
with the Lagrange multiplier $\nu$ corresponding to the normalization $\tr{\rho}=1$. Assume now that $\{\widehat{p}_i\}$ (corresponding to the state $\widehat{\rho}$) attains the maximum  over all probability distributions $\{p_i\}$ (due to continuity and compactness, this maximum is attained). We now view Eq.~(\ref{lagrangefunctionforvariance}) as a function of those variables $p_i$ for which $\widehat{p}_i>0$, in which there are $A$ such variables, and fixing the other elements $p_i$ to be zero. Then, due to the extremality of $\{\widehat{p}_i\}$ and having components in the interior of the domain of $L$, the method of Lagrange multipliers guarantees the existence of $\widehat{\nu}\in(-\infty,+\infty)$ such that
\begin{align}\label{onlytwononzerocomponentsinvariance}
\begin{split}
0~&
=~\left.\frac{dL}{dp_j}\right|_{\{\widehat{p}_i\},\widehat{\nu}}~=~(\ln\widehat{p}_j)^2+2\ln\widehat{p}_j-2\left(\sum_i\widehat{p}_i\ln\widehat{p}_i\right)(1+\ln\widehat{p}_j)+\widehat{\nu}\\
&=~\left(S(\{\widehat{p}_i\})+1+\ln\widehat{p}_j\right)^2\,-\,\left(S(\{\widehat{p}_i\})\right)^2\,+\,\widehat{\nu}-1~~~\qquad\forall j~\,\text{with}~\,\widehat{p}_j>0~,
\end{split}
\end{align}
where the quantity $S(\{\widehat{p}_i\})=S(\widehat{\rho})$ denotes the entropy of the distribution $\{\widehat{p}_i\}$ and in particular does not depend on the index $j$. Thus, the equality Eq.~\eqref{onlytwononzerocomponentsinvariance} implies that
\begin{align}
\ln\widehat{p}_j~=~\pm\sqrt{\left(S(\widehat{\rho})\right)^2-\widehat{\nu}+1}\,-\,S(\widehat{\rho})-1~~~\qquad\forall j~\,\text{with}~\,\widehat{p}_j>0~,
\end{align}
so that strict monotonicity of the logarithm yields that there can be at most two distinct non-zero elements in $\{\widehat{p}_i\}$.

Thus, leaving off hats again, an optimal $\rho=\widehat{\rho}$ has the form
\begin{align}
{\rho}~=~{\rm diag}\left(\frac{1-{r}}{m},\ldots,\frac{1-{r}}{m},\frac{{r}}{n},\ldots,\frac{{r}}{n},0,\ldots,0\right)
\end{align}
with $m+n=A$, and $r\in[0,1]$. Without loss of generality we can assume $r\leq1/2$ by permuting the entries of $\rho$. For such states, the varentropy is simple to compute,
\begin{align}\label{maximizerinproof}
(\Delta S)^2 ~=~r(1-r)\left(\ln\frac{1-r}{r}+
\ln\frac{n}{m}\right)^2~.
\end{align}
The above expression is subjected to two maximizations; the first yields that for any $r \in [0,1]$, the optimal choice is $n=A-1$ and $m=1$, whereas the second yields an optimal value $r=r_A$ satisfying
\begin{equation}
    2=(1-2r_A) \ln \frac{(1-r_A)(A-1)}{r_A}.
\end{equation}
Using the above in tandem with the inequalities
\begin{align}
x (1-x) ~ \leq ~1/4 ~ \forall x \in [0,1/2],
\end{align}
and
\begin{align}
\frac{2x(1-x)}{1-2x} \ln \frac{1-x}{x} ~ \leq ~ 1 ~ \forall x \in [0,1/2],
\end{align}
we arrive at the inequality
\begin{align}
(\Delta S)^2 ~ \leq ~ \frac{1}{4} \ln(A-1)^2 +1.
\end{align}

\end{proof}

\section{Calculations for the spin-boson model}

\setcounter{equation}{0}
\renewcommand{\theequation}{B\arabic{equation}}

In the main text, we focus on the setting of a single qubit measurement. For the sake of generality, we perform the calculation for an $m$ qubit quantum state. Recall that the environment is modeled as
\begin{equation}
    \rho^\mathcal{E} = \bigotimes_k \frac{1}{Z_k} e^{-\beta \omega_k a^\dagger_k a_k},
\end{equation}
where the subscript $k$ indicates a property or operator unique to the $k$th mode. In this generalization, each of the $m$ qubits correlate to a measurement apparatus composed of $N$ spins; the overall interaction Hamiltonian is
\begin{equation}
    H_\mathcal{AE}^\text{SB} = \sum_{i=1}^m \sum_{j=1}^N \sum_k h_{i,j} \otimes g_k (a_k^\dagger + a_k),
\end{equation}
where $h_{i,j}$ is the Hamiltonian which has the form  $\sigma_Z = \dyad*{\downarrow}-\dyad*{\uparrow}$ when acting on the $j$th spin associated with the $i$th qubit and identity on all other spins, and $g_k$ is the positive coupling constant of the $k$th environmental mode. Notice that we can write $H_\mathcal{AE}^\text{SB}=H_1 \otimes H_2$ with
\begin{equation}
    H_1 = \sum_{i=1}^m \sum_{j=1}^N h_{i,j},
\end{equation}
and
\begin{equation}
    H_2 = \sum_k g_k (a_k^\dagger + a_k).
\end{equation}

To compute the variance of $H_\mathcal{AE}^\text{SB}$, i.e
\begin{equation}
\begin{split}
    \big(\Delta H_\mathcal{AE}^\text{SB} \big)^2 &= \Tr\big(H_1^2  \rho^{\mathcal{QA}} (t) \big) \Tr\big( H_2^2 \rho^\mathcal{E}\big) - \Tr\big(H_1 \rho^{\mathcal{QA}}(t)\big)^2 \Tr\big(H_2 \rho^\mathcal{E}\big)^2,
\end{split}
\end{equation}
we make use of the fact that
\begin{equation}
    \Tr (a e^{-\beta \omega a^\dagger a}) = \Tr (a^\dagger e^{-\beta \omega a^\dagger a}) = \Tr (aa e^{-\beta \omega a^\dagger a}) = \Tr (a^\dagger a^\dagger e^{-\beta \omega a^\dagger a}) =0.
\end{equation}
This greatly simplifies the expression to
\begin{equation}
\begin{split}
    \big(\Delta H_\mathcal{AE}^\text{SB} \big)^2  &= \Tr \big( H_1^2  \rho^{\mathcal{QA}}(t) \big) \sum_{k} |g_k|^2 \Tr \big(  (a_k^\dagger a_k+a_k a_k^\dagger) \rho^{\mathcal{E}} \big) \\
    &= \Tr \big( H_1^2  \rho^{\mathcal{QA}}(t) \big)  \sum_{k} |g_k|^2 \Big( 1 +  \frac{2}{Z_k}\Tr \big( a_k^\dagger a_k e^{-\beta \omega_k a^\dagger a} \big) \Big) \\
    &= \Tr \big( H_1^2  \rho^{\mathcal{QA}}(t) \big)  \sum_{k} |g_k|^2 \Big( 1 +  \frac{2}{Z_k} \sum_{m=0}^\infty m e^{-m \beta \omega_k} \Big) \\
    &= \Tr \big( H_1^2  \rho^{\mathcal{QA}}(t) \big)   \sum_{k} |g_k|^2 \Big( 1 +  \frac{2}{e^{\beta \omega_k}-1} \Big) \\
    &= \Tr \big( H_1^2  \rho^{\mathcal{QA}}(t) \big)   \sum_{k} |g_k|^2 \coth \big( \frac{\beta \omega_k}{2} \big).
\end{split}
\end{equation}
Typically, when assuming a continuum of modes, one makes the substitution $\sum_k |g_k|^2 \rightarrow \int_0^\infty \hspace{-4pt}  J(\omega) \mathrm{d} \omega$, where $J(\omega)$ is the spectral density of the coupling constants, hence
\begin{equation}
    \big(\Delta H_\mathcal{AE}^\text{SB} \big)^2 = \Tr \big(H_1^2  \rho^{\mathcal{QA}}(t)\big) \int_0^\infty \hspace{-7pt} J (\omega) \coth  \big( \frac{\beta \omega}{2} \big) \mathrm{d} \omega.
\end{equation}
Finally, recall that all of the spins assigned to a qubit are correlated with each other, thus for any $j_1$ and $j_2$
\begin{equation}
    \Tr \big( h_{i_1,j_1} h_{i_2,j_2} \rho^{\mathcal{QA}} (t) \big) = \Tr \big( h_{i_1,1} h_{i_2,1} \rho^{\mathcal{QA}} (t) \big).
\end{equation}
Using this innate symmetry with the spins of the apparatus, one obtains the expression 
\begin{equation}
     \big(\Delta H_\mathcal{AE}^\text{SB} \big)^2 = \chi N^2 \int_0^\infty \hspace{-7pt} J (\omega) \coth  \big( \frac{\beta \omega}{2} \big) \mathrm{d} \omega,
\end{equation}
where
\begin{equation}
\label{eq:pre-factorChi}
    \chi = \sum_{i_1,i_2=1}^m \Tr \big( h_{i_1,1}h_{i_2,1}  \rho^{\mathcal{QA}}(t)\big)
\end{equation}
is a time-independent pre-factor dependent on the initialization of the quantum state; note that $\chi=1$ for a single qubit measurement ($m=1$). Therefore, the timescale of a measurement can be bounded via
\begin{equation}
    \tau \geq \frac{\hbar \; \delta S}{2N \sqrt{\chi}} \Big( \int_0^\infty \hspace{-7pt} J(\omega) \coth(\beta \omega /2) \mathrm{d} \omega \Big)^{-1/2}.
\end{equation}

In the single-qubit case, a tighter bound can be obtained by employing the Born-Markov approximation and maximizing $\Delta S \sqrt{I_d}$. As stated in the main text, the off-diagonal terms of $\rho^\mathcal{QA}(t)$ acquire a decay term $e^{-\Gamma}$, with
\begin{equation}
    \label{eq:Gamma_approx}
    \Gamma = 4 N \int_0^\infty \hspace{-4pt} \frac{J(\omega)}{\omega^2}(1-\cos (\omega t/\hbar) ) \coth (\beta \omega /2) \mathrm{d} \omega \approx \frac{2 N t^2}{\hbar^2} \int_0^\infty \hspace{-4pt} J(\omega) \coth (\beta \omega /2) \mathrm{d} \omega,
\end{equation}
thus if the state after the pre-measurement is
\begin{equation}
    \ket*{\psi^\mathcal{QA}} = x \ket*{0}\ket*{ \downarrow }^{\otimes N} + y \ket*{1} \ket*{\uparrow}^{\otimes N},
\end{equation}
the eigenvalues of the decohering state are
\begin{equation}
    \lambda_\pm = \frac{1}{2} \Big( 1 \pm \sqrt{1-4|xy|^2(1-e^{-2\Gamma})} \Big).
\end{equation}
It is straightforward to compute
\begin{equation}
    (\Delta S)^2 = \lambda_+ (\ln \lambda_+)^2 + \lambda_- (\ln \lambda_-)^2 - \big( \lambda_+ \ln \lambda_+ + \lambda_- \ln \lambda_- \big)^2 = \lambda_+ \lambda_- \ln \left( \frac{\lambda_+}{\lambda_-} \right)^2,
\end{equation}
and therefore
\begin{equation}
    I_d = \frac{1}{\lambda_+} \left( \frac{\partial \lambda_+}{\partial t}\right)^2 + \frac{1}{\lambda_-} \left( \frac{\partial \lambda_-}{\partial t}\right)^2 = \frac{1}{\lambda_+ \lambda_-}\frac{16|xy|^4}{(\lambda_+-\lambda_-)^2} \frac{\Gamma^2}{t^2}e^{-4\Gamma} = \frac{16|xy|^2 \Gamma}{t^2} \frac{1}{(\lambda_+-\lambda_-)^2} \frac{\Gamma e^{-4\Gamma}}{1-e^{-2\Gamma}},
\end{equation}
where we use the approximation in Eq.~\eqref{eq:Gamma_approx} and note that $\Gamma/t^2$ is thus time-independent. By combining the above two equations, one obtains
\begin{equation}
    (\Delta S)^2 I_d = \frac{8|xy|^2 \Gamma}{t^2} \Big(\sqrt{\lambda_+ \lambda_-}\frac{\ln \lambda_+-\ln \lambda_-}{\lambda_+-\lambda_-}  \Big)^2 \left( \frac{2\Gamma e^{-4\Gamma}}{1-e^{-2\Gamma}} \right),
\end{equation}
which can be bounded above by employing the inequalities
\begin{equation}
    \left(\sqrt{\lambda_+ \lambda_-}\frac{\ln \lambda_+-\ln \lambda_-}{\lambda_+-\lambda_-}  \right)^2 \leq 1,
\end{equation}
and
\begin{equation}
    \frac{2\Gamma e^{-4\Gamma}}{1-e^{-2\Gamma}} \leq 1.
\end{equation}
Therefore,
\begin{equation}
    \Delta S \sqrt{I_d} \leq \sqrt{\frac{8|xy|^2\Gamma}{t^2}} \leq \sqrt{\frac{2\Gamma}{t^2}} = \frac{2}{\hbar}\sqrt{N} \Big( \int_0^\infty \hspace{-7pt} J(\omega) \coth(\beta \omega /2) \mathrm{d} \omega \Big)^{1/2}
\end{equation}
from which it follows that under the Born-Markov approximation, the measurement time of a single qubit is bounded by
\begin{equation}
    \tau \geq \frac{\hbar \; \delta S}{2 \sqrt{N}} \Big( \int_0^\infty \hspace{-7pt} J(\omega) \coth(\beta \omega /2) \mathrm{d} \omega \Big)^{-1/2}.
\end{equation}

\section{Calculations for the boson-boson model}

\setcounter{equation}{0}
\renewcommand{\theequation}{C\arabic{equation}}

The boson-boson example calculations are very similar to the spin-boson calculations. For a multi-qubit example, we could consider $m$ qubits, each coupled to a measurement apparatus with outputs $\ket*{\alpha}$ and $\ket*{-\alpha}$. This generalization leads to a similar pre-factor of $\chi$, Eq.~\eqref{eq:pre-factorChi}. Thus, for compactness, we only consider a single qubit in this set of calculations.

Using
\begin{equation}
     H_\mathcal{AE}^\text{BB} = \sum_k g_k (b a^\dagger_k +  b^\dagger a_k),
\end{equation}
the only non-vanishing terms in the variance are
\begin{equation}
    \big(\Delta H_\mathcal{AE}^\text{BB} \big)^2 = \sum_k g^2_k \Tr \big( (b b^\dagger)(a^\dagger_k a_k)+ (b^\dagger b)(a_k a^\dagger_k) \rho^\mathcal{QAE} \big).
\end{equation}
From $b \ket*{\alpha}=\alpha \ket*{\alpha}$, it follows that
\begin{equation}
\begin{split}
    \big(\Delta H_\mathcal{AE}^\text{BB} \big)^2 &= \sum_k g^2_k \Big( (1+|\alpha|^2) \Tr \big(a^\dagger_k a_k \rho^\mathcal{E} \big)+ |\alpha|^2 \Tr \big(a_k a_k^\dagger \rho^\mathcal{E} \big) \Big) \\
    &\approx | \alpha|^2 \sum_k g_k^2  \Tr \big((a^\dagger_k a_k+a_k a_k^\dagger) \rho^\mathcal{E} \big) \\
    &= |\alpha|^2 \sum_k g^2_k \coth(\frac{\beta \omega_k}{2})\\
    &\rightarrow |\alpha|^2 \int_0^\infty \hspace{-7pt} J(\omega) \coth(\frac{\beta \omega}{2}) \mathrm{d}\omega,
\end{split}
\end{equation}
where we assume that $|\alpha|^2 \gg 1$, hence $1+|\alpha|^2 \approx |\alpha|^2$.

\end{document}